\documentclass[a4paper]{VisionStyle}
\usepackage{epsfig}

\begin{document}

\title{First Results from a \emph{XMM--Newton} Survey of a Distance-Limited ($D < 22$~Mpc) Sample
of Seyfert Galaxies. II -- The Galactic Serendipitous Sources}

\author{L. Foschini\inst{1}, G. Di Cocco\inst{1}, M. Dadina\inst{1}, L. Bassani\inst{1}, M. Cappi\inst{1},
J. B. Stephen\inst{1}, M. Trifoglio\inst{1}, F. Gianotti\inst{1}, L. C. Ho\inst{2}, J. Mulchaey\inst{2},
F. Panessa\inst{1,3}, E. Piconcelli\inst{1,3}}

\institute{
	IASF -- CNR, Sezione di Bologna (formerly iTeSRE). Via Gobetti 101, 40129 Bologna (Italy)
\and
	The Observatories of the Carnegie Institution of Washington, 813 Santa Barbara Street, Pasadena, CA 91101 (USA)
\and
	Dipartimento di Astronomia, Universit\`a di Bologna (Italy)}

\maketitle

\begin{abstract}
We present preliminary results from a study with \emph{XMM--Newton} of the galactic
serendipitous sources in a sample of 28 nearby Seyfert galaxies (with $D < 22$ Mpc).
We concentrate here on a subset of four galaxies, namely: NGC3185, NGC3486, NGC3941, and
NGC4565. In these galaxies we found six ultraluminous X--ray source (ULX). The
large collecting area of \emph{XMM--Newton} makes the statistics sufficient to perform
spectral fitting with simple models. We shortly discuss the results obtained so far.
\end{abstract}

\section{Introduction}
The X--ray emission from Seyfert host galaxies is essentially composed by the contribution of a number of discrete
sources and the hot interstellar plasma (Fabbiano 1989).
Most of the discrete sources appear to be close accreting binaries, with a compact companion.
\emph{Einstein} observations of the bulge of M31 showed a population of about 100 low mass X--ray binaries LMXRB
(Fabbiano et al. 1987). Later on, Supper et al. (1997) showed, by using \emph{ROSAT} data, that the most
luminous of these objects has $L_X=2\times 10^{38}$ erg/s, close to the Eddington limit for a $1.4 M_{\sun}$
neutron star .

Recently, several sources with X--ray luminosity quite higher than the Eddington limit for a neutron star have
been detected in nearby galaxies (e.g. Read et al. 1997, Makishima et al. 2000, Zezas et al. 2001).
The highest luminosity value is about $2\times 10^{40}$ erg/s. Fabbiano et al. (2001) found with
\emph{Chandra} 14 pointlike sources in the Antennae galaxies, with luminosities above $10^{39}$ erg/s
and up to $10^{40}$ erg/s.

This discovery has raised difficulties to interpret these sources. Even though it is statistically possible
to have \emph{some} individual cases of off--centre black holes with masses of the order of
$10^{3}-10^{4}$ $M_{\sun}$ (by assuming a typical Eddington ratio of $0.1-0.01$; cf. Nowak 1995),
it is very difficult to explain the high number of sources detected so far within this scenario.
Several other hypotheses have been suggested about the nature of ULX: anisotropic emission from accreting black holes
(King et al. 2001), emission from jets in microblazar (K\"ording et al. 2002), emission from accreting Kerr
black holes (Makishima et al. 2000).
However, the lack of sufficient spectral information did not allow to date to distinguish among the different
models proposed.

Here we present early results of the analysis of six of the most luminous galactic serendipitous sources detected
with \emph{XMM--Newton} during a survey of a distance--limited ($D < 22$ Mpc) sample of Seyfert galaxies (Di Cocco et al. 2000,
Cappi et al. these Proceedings).

\begin{table*}[!ht]
\caption{Coordinates of the selected sources in the galaxies observed with XMM--Newton. ROSAT detections are
taken from Vogler et al. (1996). The Distance column indicates the angular distance from the optical centre of
the galaxy in arcsec and in kpc between brackets.}
\centering
\begin{tabular}{lcclc}
\hline
Object & Host Galaxy & RA, Dec (J2000)  & Distance & ROSAT Position\\
\hline
XMMU J101737.4+214145 & NGC3185     & 10:17:37.4, +21:41:45           & 33 (3.4)     & --\\
XMMU J110022.4+285818 & NGC3486     & 11:00:22.4, +28:58:18           & 26 (0.9)     & --\\
XMMU J115258.2+365858 & NGC3941     & 11:52:58.2, +36:58:58           & 36 (3.3)     & --\\
XMM J123617.5+285855  & NGC4565     & 12:36:17.5, +25:58:55           & 52 (2.4)     & RX J1236.2+2558\\
XMM J123614.6+260051  & NGC4565     & 12:36:14.6, +26:00:51           & 131 (6.2)    & RX J1236.2+2600\\
XMM J123627.5+255733  & NGC4565     & 12:36:27.5, +25:57:33           & 133 (6.3)    & RX J1236.4+2557\\
\hline
\end{tabular}
\label{tab:srccoord}
\end{table*}

\begin{figure*}[!ht]
  \begin{center}
    \epsfig{file=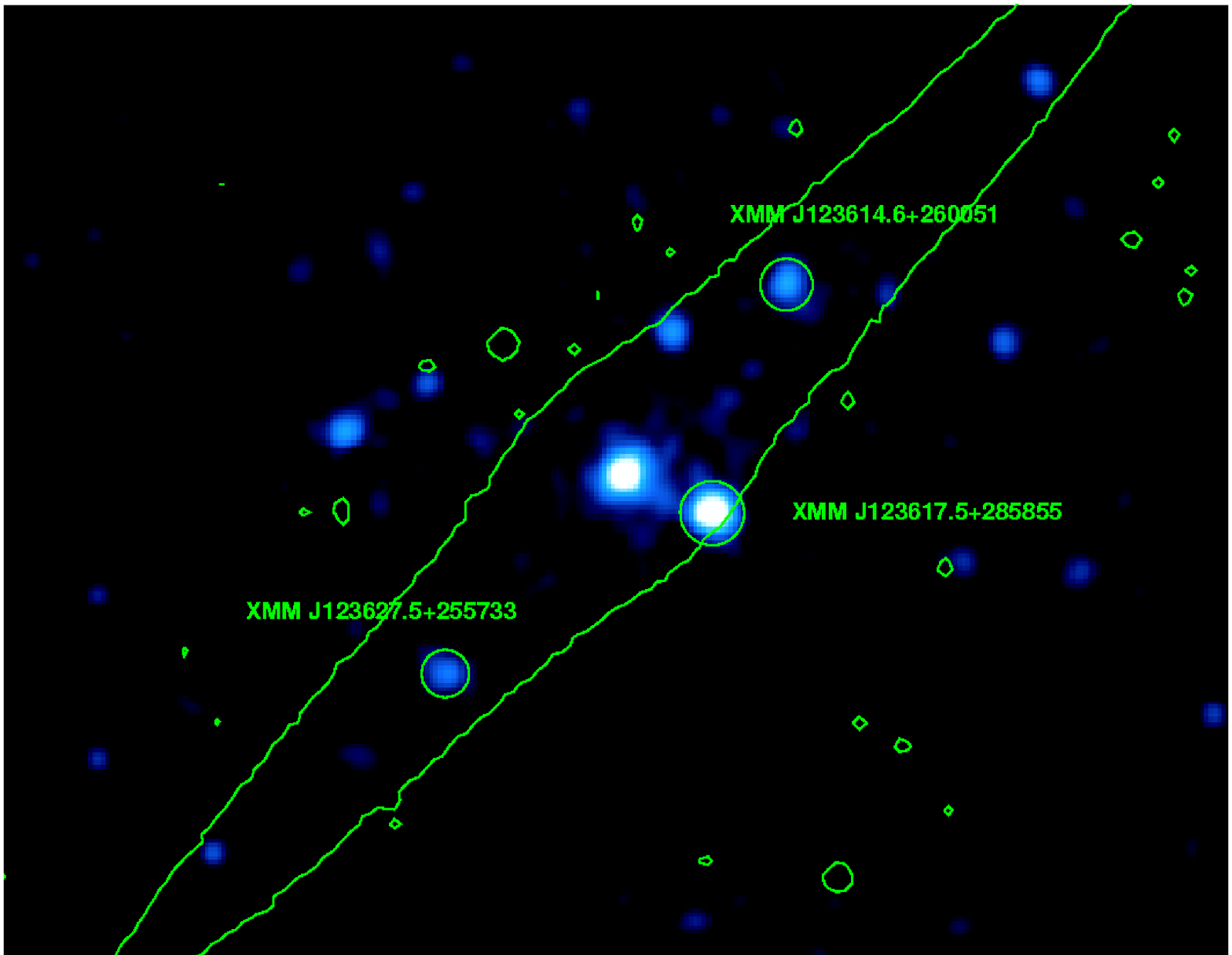, angle=270, width=8cm}
    \hspace{12pt}
    \epsfig{file=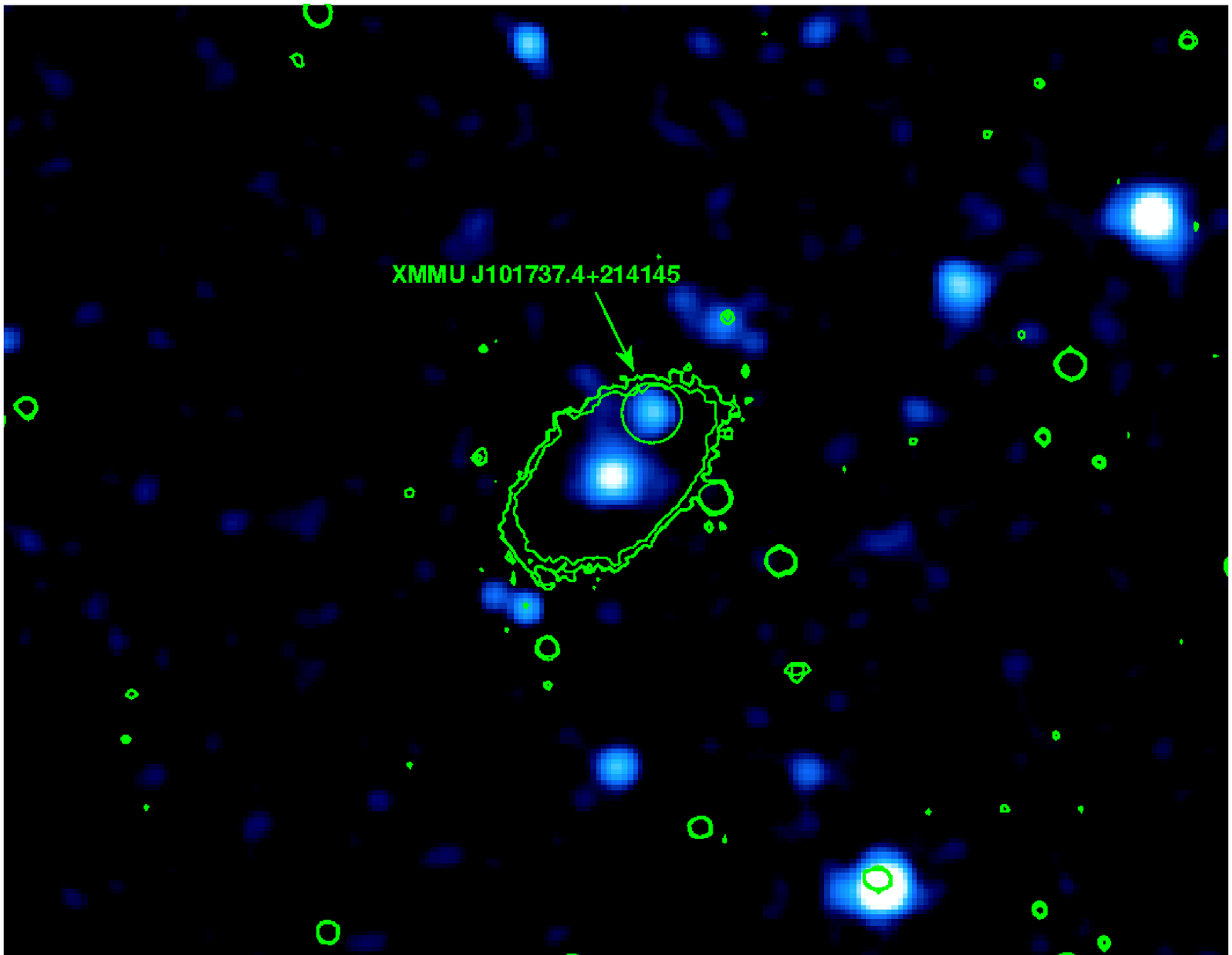, angle=270, width=8cm}
  \end{center}
\caption{Images from MOS1 data, in the $0.2-10$~keV energy band, with contours from optical images (DSS)
superposed. (A, left) NGC4565, with the three sources encircled (cf. Table\ref{tab:srccoord}); (B, right) NGC3185 with XMMU J101737.4+214145.}
\label{fig1}
\end{figure*}

\section{XMM--Newton data}
The present analysis refers to the six brightest off--nuclear sources to date detected in our sample
of nearby galaxies (Table~\ref{tab:srccoord}). The nominal exposures are about 10 ks for each target.
Proton flares reduced the useful exposure of NGC3486 and NGC3941
observations to 5 and 6.5 ks, respectively.

\begin{figure*}[ht]
  \begin{center}
    \epsfig{file=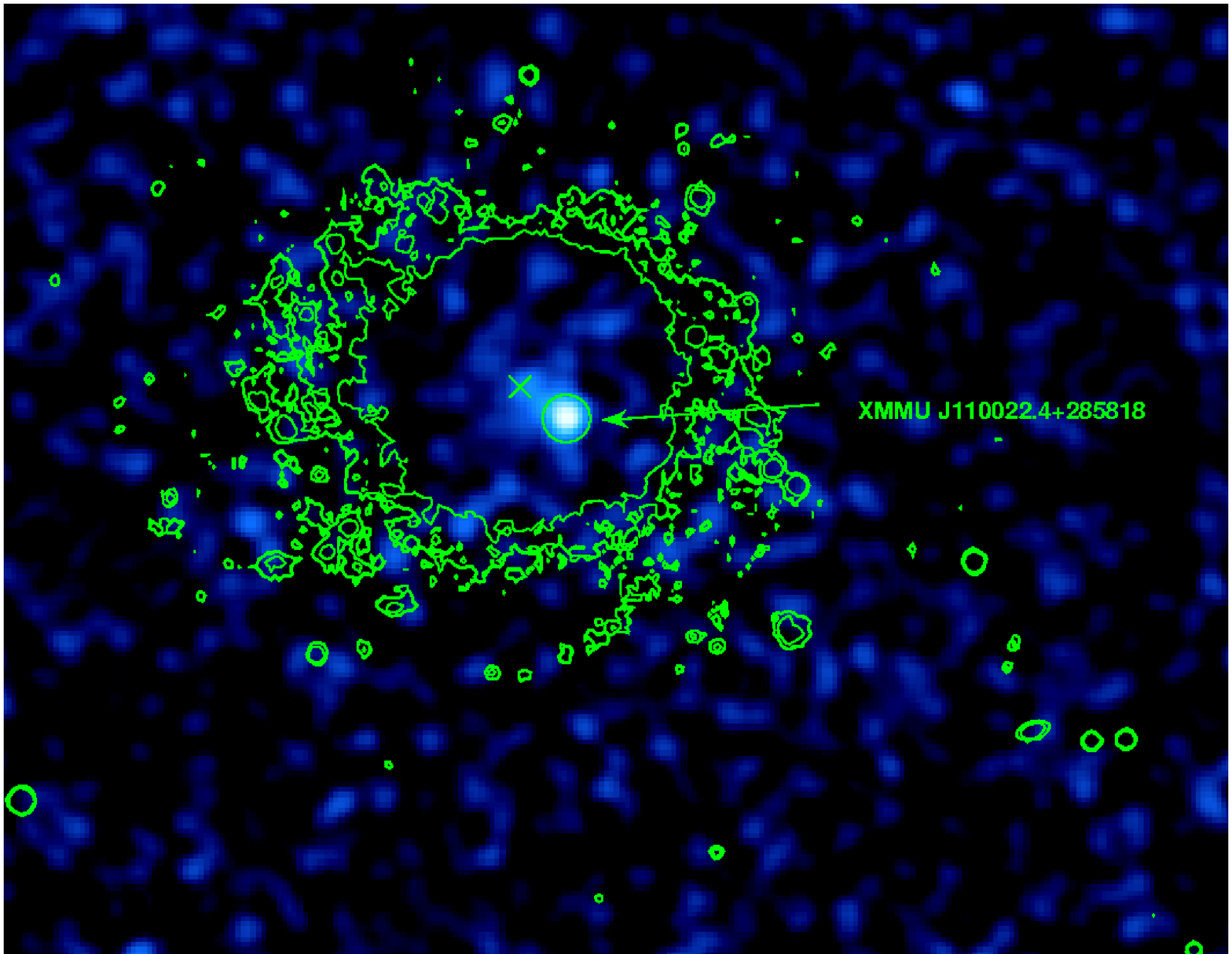, angle=270, width=8cm}
    \hspace{12pt}
    \epsfig{file=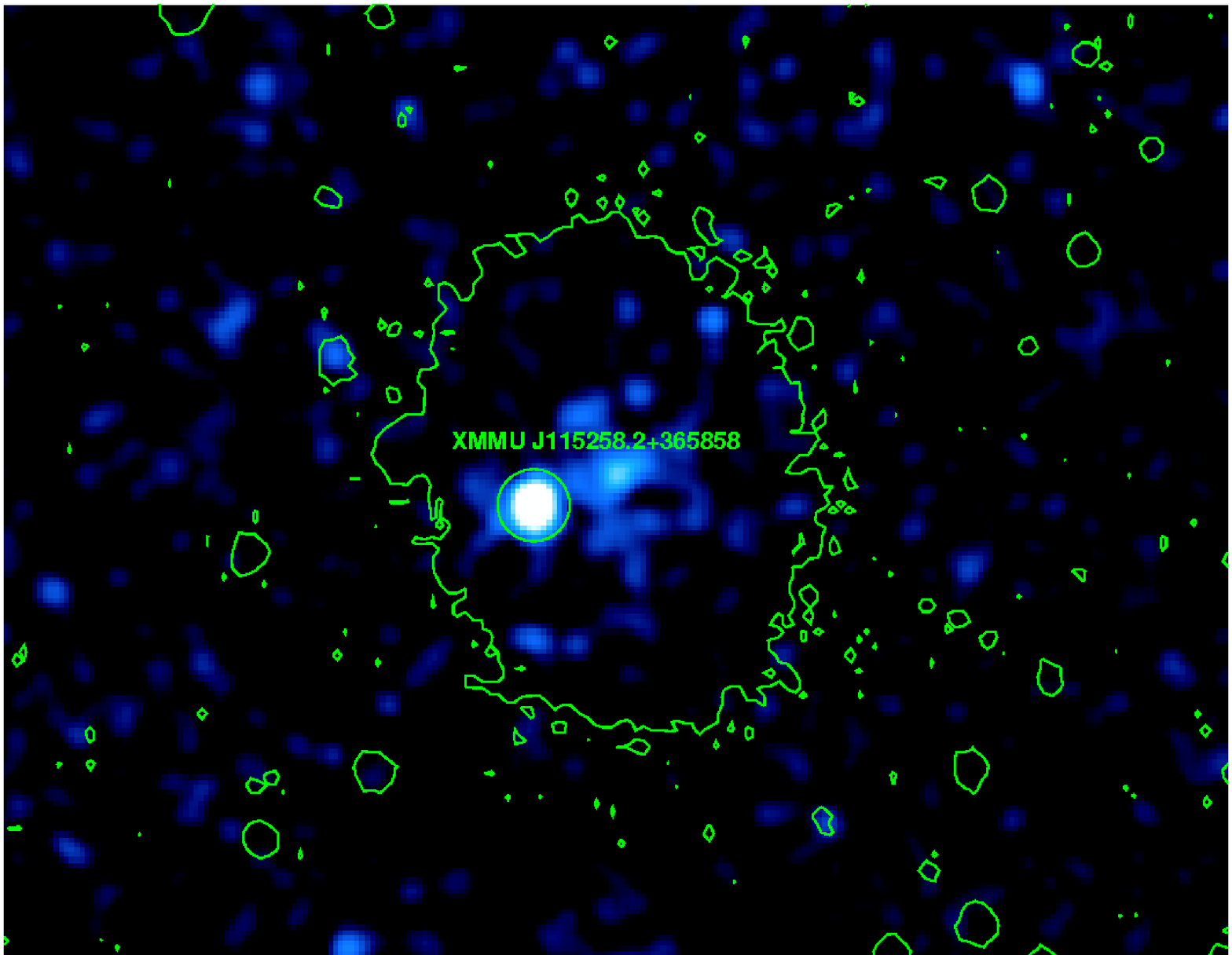, angle=270, width=8cm}
  \end{center}
\caption{Images from MOS1 data, in the $0.2-10$~keV energy band, with contours from optical images (DSS) superposed. Refer to
Table~\ref{tab:srccoord} for more details on the source locations. (A, left) NGC3486, the cross indicate the optical centre of
the host galaxy; (B, right) NGC3941. In both cases, the off--nuclear source is more luminous than the nucleus.}
\label{fig2}
\end{figure*}

To define an X--ray source as an off--nuclear galactic object, we used two selection criteria:
the object has to be inside the DSS (Digitized Sky Survey) optical contours of the host galaxy
and sufficiently far from the optical centre, to avoid confusion
with galaxy active nucleus.
Considering an absolute location accuracy for \emph{XMM--Newton} of about $4''$ (Jansen et al. 2001),
and combining it with the uncertainty of $2.5''$ in the optical position, we searched for off--nuclear sources
at least $10''$ away from the optical centre of the host galaxy. Among the several detections in the
sample available to date, we select the six brightest sources (see Fig.~\ref{fig1}, Fig.~\ref{fig2}),
three of which are still unidentified, while the remaining three were already detected with ROSAT (Vogler et al. 1996).
The coordinates and the distances from the optical centre of the galaxy are summarized in Table~\ref{tab:srccoord}.

For the processing, screening, and analysis of the data we used the standard tools
of XMM--SAS software v. 5.0.1 and HEAsoft Xspec (11.0.1). The images are prepared
with DS9 v. 2.1, together with ZHTools v. 2.0.2. We do not apply the correction for
vignetting, because all the selected sources are close to the centre of the FOV
(less than $2'$).

\section{Spectral fitting}
Statistics are generally good enough to perform a spectral fitting with simple models. After a survey of
existing literature (cf., for example, Colbert \& Mushotzky 1999, Makishima et al. 2000) we decided to
fit the data from MOS1, MOS2, and PN, with the following models:
power law (PL), black body (BB), thermal bremstrahlung (BR), unsaturated Comptonization (CST) by
Sunyaev \& Titarchuk (1980), and the multicolor black body accretion disk (MCD) by Mitsuda et al. (1984).
The results are summarized in Table~\ref{tab:sorg2}, where we report only the fits with
reduced--$\chi^2$ lower than 2.

\begin{table*}[t]
\caption{Summary of the spectral fitting based on combined MOS1, MOS2, and PN data for the six sources.
Uncertainties in the parameters estimate are at the $90\%$ confidence limits. Distances are taken from Ho et al. (1997).}
\centering
\begin{tabular}{llcccc}
\hline
Model$^{\mathrm{a}}$ & Parameter   & $N_H$$^{\mathrm{b}}$  & $\chi^2/\nu$      & Flux 0.5-10 keV                       & Lumin. 0.5-10 keV \\
{}                   & {}          & [$10^{22}$ cm$^{-2}$] & {}                & [$10^{-14}$ erg cm$^{-2}$ s$^{-1}$]   & [$10^{39}$ erg/s]\\
\hline
\multicolumn{1}{c}{Source:}&{XMMU J101737.4+214145}&{(NGC3185)}&{($D=21.3$ Mpc)}&{}&{}\\
\hline
PL     & $\Gamma = 1.7 \pm 0.8$ & --                      & 6.3/6             & 3.1                           & 1.7\\
BB     & $kT=0.4\pm 0.2$ keV    & --                      & 10.3/6            & 1.4                           & 0.8\\
\hline
\multicolumn{1}{c}{Source:}&{XMMU J110022.4+285818}&{(NGC3486)}&{($D=7.4$ Mpc)}&{}&{}\\
\hline
PL     & $\Gamma = 2.2 \pm 0.5$ & --                      & 7.7/14            & 8.3                           & 0.5\\
BB     & $kT=0.26\pm 0.05$ keV  & --                      & 13.3/14           & 4.3                           & 0.3\\
MCD    & $kT=0.4\pm 0.2$ keV     & --                      & 11.9/14           & 4.9                           & 0.3\\
\hline
\multicolumn{1}{c}{Source:}&{XMMU J115258.2+365858}&{(NGC3941)}&{($D=18.9$ Mpc)}&{}&{}\\
\hline
PL     & $\Gamma = 1.9 \pm 0.2$ & --                      & 17.8/19           & 17.1                          & 7.4\\
BR     & $kT = 4 \pm 2$ keV     & --                      & 24.4/19           & 14.4                          & 6.2\\
\hline
\multicolumn{1}{c}{Source:}&{XMM J123617.5+285855}&{(NGC4565)}&{($D=9.7$ Mpc)}&{}&{}\\
\hline
PL     & $\Gamma = 2.0 \pm 0.2$ & $< 0.06$                & 42.7/45           & 20.3                           & 2.4\\
BR     & $kT = 3.4 \pm 0.8$ keV & --                      & 48.4/46           & 17.8                           & 2.0\\
CST    & $kT = 1.8 \pm 0.6$ keV & --                      & 42.8/45           & 19.4                           & 2.1\\
{}     & $\tau_e = 16\pm 7$     & {}                      & {}                & {}                             & {}\\
MCD    & $kT = 0.8 \pm 0.1$ keV  & --                      & 83.1/46           & 14.2                           & 1.6\\
\hline
\multicolumn{1}{c}{Source:}&{XMM J123614.6+260051}&{(NGC4565)}&{($D=9.7$ Mpc)}&{}&{}\\
\hline
PL     & $\Gamma = 1.8 \pm 0.5$ & $0.6\pm0.4$             & 17.1/12           & 10.6                            & 1.7\\
BB     & $kT=0.9\pm 0.1$ keV    & --                      & 17.7/13           & 7.5                             & 0.8\\
MCD    & $kT=2\pm 1$ keV         & --                      & 20.8/13           & 10.4                            & 1.1\\
\hline
\multicolumn{1}{c}{Source:}&{XMM J123627.5+255733}&{(NGC4565)}&{($D=9.7$ Mpc)}&{}&{}\\
\hline
PL     & $\Gamma = 2.7 \pm 0.9$ & $0.5\pm 0.4$            & 13.5/9            & 5.1                           & 1.2\\
BB     & $kT = 0.52 \pm 0.07$ keV  & --                      & 13.5/10           & 3.9                           & 0.4\\
MCD    & $kT = 1.0\pm 0.3$ keV      & --                      & 15.1/10           & 4.7                           & 0.5\\
\hline
\end{tabular}
\label{tab:sorg2}
\begin{list}{}{}
\item[$^{\mathrm{a}}$] PL: power law; BB: black body; BR: thermal bremsstrahlung; CST: unsaturated Comptonization; MCD: multicolor
accretion disk.
\item[$^{\mathrm{b}}$] Intrinsic absorption of the source, in addition to the galactic $N_H$.
\end{list}
\end{table*}

\begin{figure*}[!ht]
  \begin{center}
    \epsfig{file=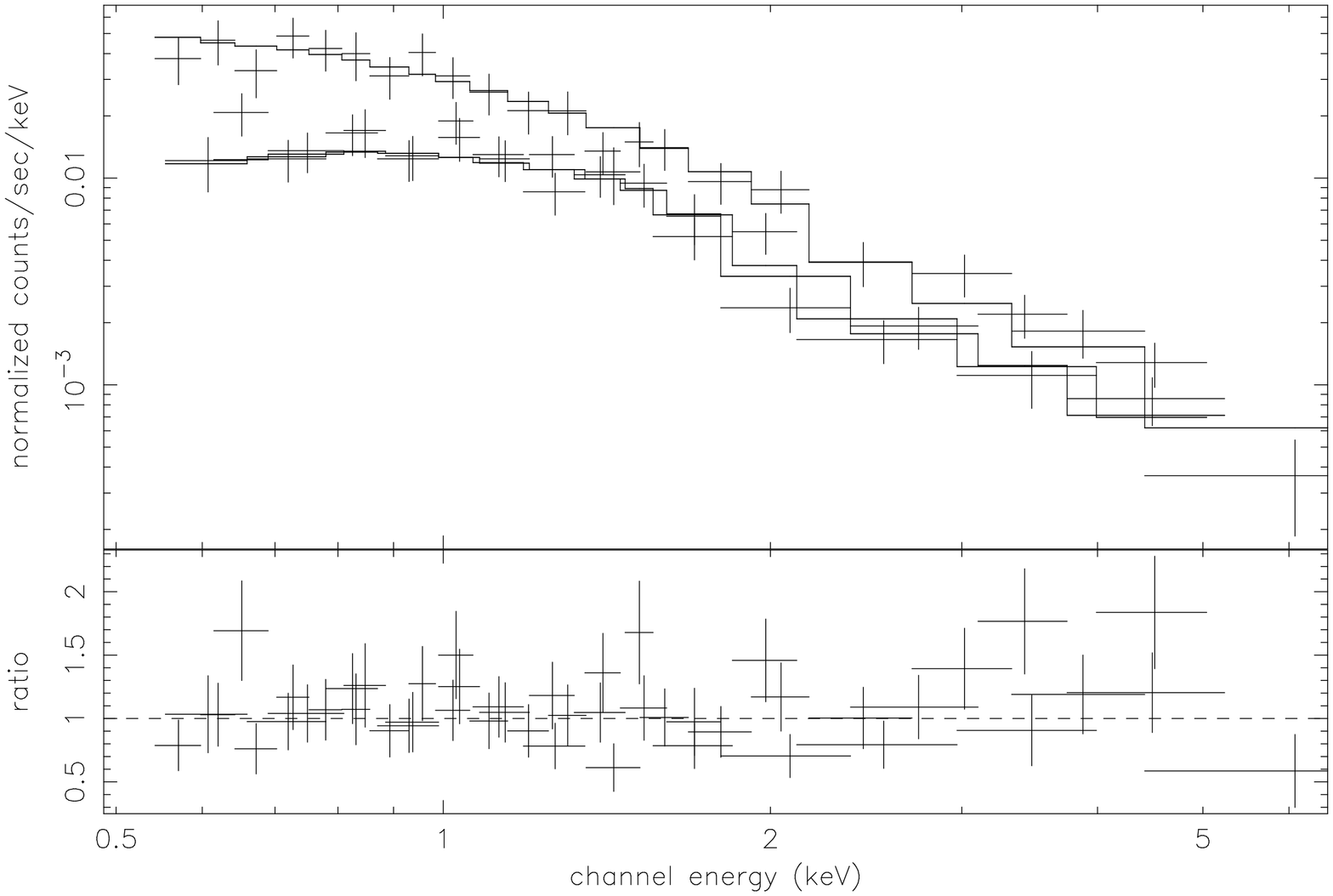, angle=270, width=7.5cm}
    \hspace{12pt}
    \epsfig{file=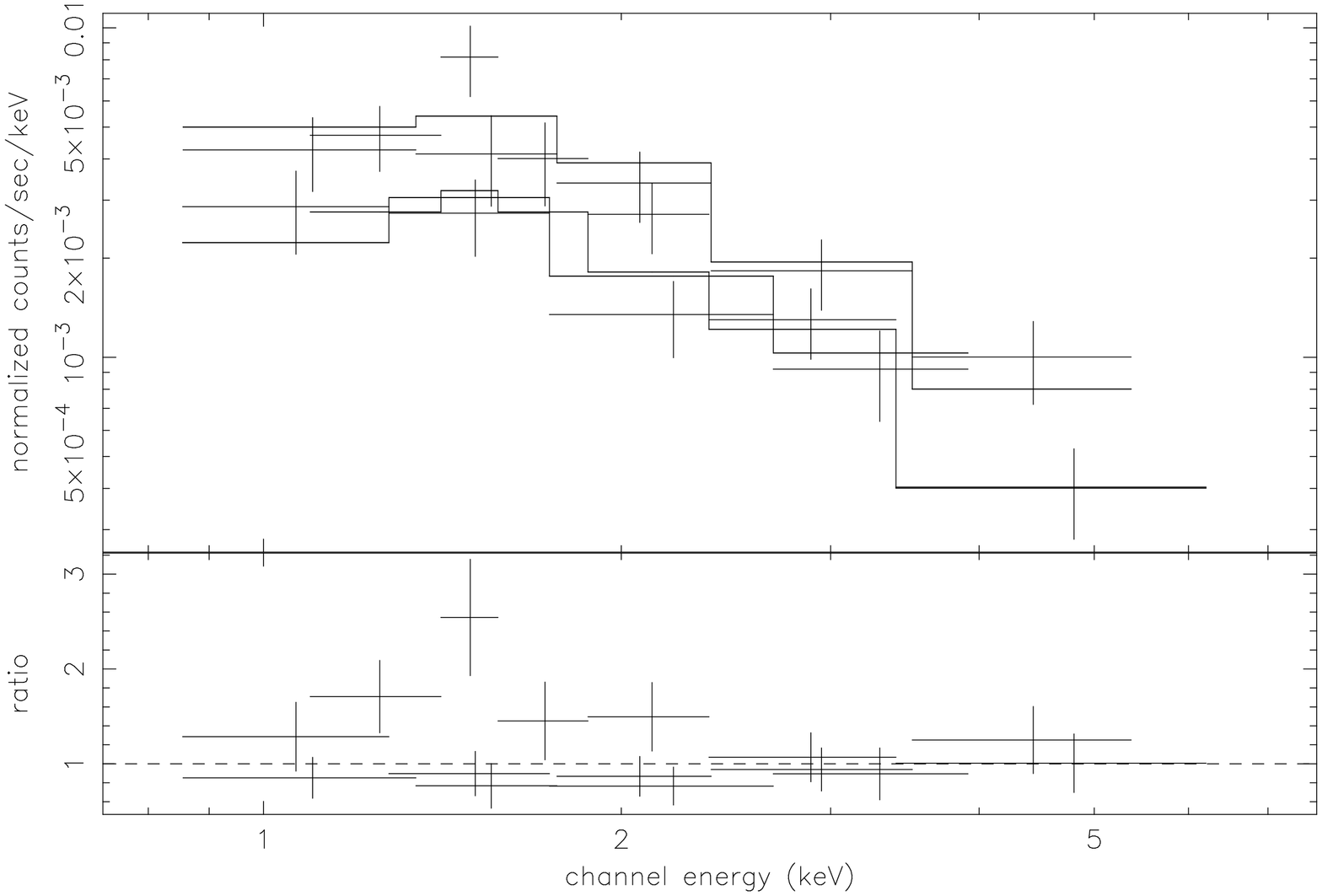, angle=270, width=7.5cm}
  \end{center}
\caption{Combined spectrum (MOS1, MOS2, PN) of (A, left) the source XMM J123617.5+285855 in NGC4565, fitted with a
power law model and (B) the source XMM J123614.6+260051, fitted with a black body model.
Refer to Table 2 for more details.}
\label{fig3}
\end{figure*}

For the first four sources in Table~\ref{tab:sorg2} provides the best fit with a simple power law
with $\Gamma \approx 1.7-2.2$ (for an example of spectrum, see Fig.~\ref{fig3}A). The two remaining
cases are best fitted with a black body model, with $kT\approx 0.5-0.9$~keV (for an example see Fig.~\ref{fig3}B).
It is known that the emission expected from a black hole X--ray binary (BHXRB) is variable:
in the hard state, the spectrum is typically a power law with $\Gamma \approx 1.3-1.9$,
while in the soft state the spectral index increases up to about 2.5 and a soft component appears in
the X--ray spectrum (e.g. Ebisawa et al. 1996). Therefore, given the actual statistics, our
sources could be BHXRB in a hard/soft state.
The exposures are too short to perform meaningful time variability studies. This is unfortunate
because variability could help to distinguish XRB from, e.g., young X--ray supernovae remnants.

It is interesting to note that the MCD model, which has often been successfull in the past for ULX (e.g.
Colbert \& Mushotzky 1999, Makishima et al. 2000), does not fit well most of our sources.
Even when we obtain a reasonable fit with MCD (for the last two sources
in Table~\ref{tab:sorg2}), statistics give a better fit with a simple black body model.

The unsaturated Comptonization (CST) does not give acceptable fits, except for XMM J123617.5+285855, for which
it represents the second best fit, after the power law (with $\Gamma = 2.0\pm 0.2$).

\section{Luminosities}
The luminosities observed are in the range $4 - 65\times 10^{38}$ erg/s, depending on the model considered. Under the assumption of
uniform and isotropic accretion, a bolometric luminosity approximately equal to the X--ray luminosity, and an Eddington
ratio of $1$, these values suggest the presence of compact objects with masses between $3$ and $50$ $M_{\sun}$.
If the Eddington ratio is in the range of about $0.1-0.01$, as suggested by observations of galactic black holes
(e.g. Nowak 1995), this would imply a mass range of $30-5000$ $M_{\sun}$.

Such high masses are difficult to explain for off--centre sources and several authors
have proposed alternatives to the above scenario. For example,
Makishima et al. (2000) proposed a Kerr BH scenario: in this case, the luminosity produced by a spinning BH can be
up to 7 times larger than in a Schwarzschild BH. On the other hand, King et al. (2001) suggested that the matter could accrete
anisotropically: an anisotropic factor of $0.1-0.01$ reduces the values of the mass to those typically observed in
X--ray binaries in the Milky Way. Finally, K\"ording et al. (2002) suggested the possibility of relativistic beaming
due to the presence of jets coupled to an accretion disk. With respect to this scenario, it is
worth noting that Mirabel \& Rodriguez (1999) found that the speed of the relativistic jet has a bimodal distribution. For
an accreting neutron star $v\approx 0.3c$, lower that the value related to the case of an accreting black hole ($v\approx 0.9c$).
In this scenario, if the angle between the line of sight and the axis of ejection is very small (less than $10\degr$),
the flux density is boosted up to $8\gamma^3$ ($\gamma$ is the Lorentz factor)
with respect to its value in the rest frame. With the speed values cited above, the flux boosting
for an accreting neutron star could be increased by about 10 times, while it could be even 100 times if a black
hole is present (cf. Georganopoulos et al. 2001).

\section{Final remarks}
We present preliminary results from the analysis of the brightest sources detected in a sample of four nearby spiral galaxies.
Statistics are sufficient to perform a spectral fitting with simple models. What is emerging from this preliminary
analysis is that their spectral properties, given the present statistics, are best consistent with models
typical of accreting BH binaries in a hard/soft state. But the luminosities observed probably require some
anisotropy or beaming in the emission. Observations with higher exposures and at other wavelengths (particularly
optical and radio) are required to put tighter constraints on the nature of these enigmatic objects.

\begin{acknowledgements}
This work is based on observations obtained with \emph{XMM--Newton}, an ESA science
mission with instruments and contributions directly funded by ESA Member States and the USA (NASA).
We gratefully acknowledge partial support funding from the Italian Space Agency (ASI).
This research has made use of \emph{NASA's Astrophysics Data System Abstract Service}.
\end{acknowledgements}

\end{document}